\documentstyle[twocolumn,aps]{revtex}

\begin{document}

{\bf Comment on ``Gravity Waves, Chaos, and Spinning Compact
Binaries''}

\medskip

Levin has shown that spinning compact binaries can be chaotic at second
post-Newtonian order. However, when higher order dissipational effects
are included, the dynamics will no longer be chaotic, though the evolution
may still be unpredictable in a practical sense. I discuss some of
the additional work that needs to be done to decide how this
unpredictability might affect gravitational
wave detectors such as LIGO.

Levin\cite{janna} has established that the
two body problem, with spin, is chaotic at second Post-Newtonian (2PN)
order. This important result supports and
extends Suzuki and Maeda's\cite{sm} demonstration that the system is
chaotic in the test particle limit. The 2PN equations of motion are
conservative, and the system is chaotic in a strict mathematical sense.
Indeed, the full two body problem conserves total energy\cite{ash}, but
it is natural to treat the flux of gravitational waves as an energy loss
mechanism, and the orbital dynamics as dissipative. The first dissipative
terms in the orbital dynamics show up at 2.5PN order. To be precise, the
orbital dynamics is damped and weakly driven, since some of the
gravitational waves backscatter off the spacetime curvature and
are re-absorbed by the orbiting bodies. However, the energy loss
always exceeds the energy re-absorbed. Consequently, the phase
space has a simple attracting fixed point, and is not formally chaotic.

The final section of Levin's paper deals with the effect of dissipation
on the system. Levin notes that ``some orbits will sweep through the
chaotic region of phase space as they inspiral'' and that ``dissipation
does not obliterate the chaos''. An apparently fractal basis boundary is
displayed in Fig.~5 to support this point. However, since there are no
unstable quasi-periodic
orbits in a damped system, the boundaries can not be truly fractal.
If we were to ``zoom in'' on a small portion of the boundary we would see
that it is actually smooth. Thus, in a strict mathematical sense, the
boundaries are not fractal and the dynamics is not chaotic. I believe that
Levin is using the term ``fractal scaling'' in the way it is used to
describe real objects such as trees or coastlines.
These scaling laws only hold over a finite
range of length scales. For the two body problem it would be interesting
to study how the number of decades of fractal scaling varies with the strength
of the dissipation. The larger the range of effectively fractal scaling,
the more unpredictable the system is in a practical sense.

The situation here is reminiscent of the problems encountered when studying the
quantum analogs of classically chaotic systems and the recovery of chaos in
the classical limit. It was argued that quantum systems
could not be chaotic since the uncertainty principle does not allow there to be
fractal structures in phase space. Nevertheless, a shadowy imprint of the
classical fractal structure was found to remain\cite{gutz1}, and it is now
recognized that energy levels in the quantum analogs of classically integrable
systems differ from those in classically chaotic systems\cite{rmt}. Perhaps
something similar applies here. As Frankel and I argued earlier\cite{us}, the
dissipative dynamics can be approximated by a sequence of conservative
orbits with decreasing energy and angular momentum. The orbits of the
dissipative system are effectively sewn together from the threads of a
conservative chaotic system. Consequently, the behaviour of the dissipative
system can be erratic and difficult to predict,
even though the dynamics is not chaotic.

The most important questions still to be addressed are how prevalent and how
strong the erratic behaviour might be, and what effect this might have on
the detection of graviational waves. The question of prevalence can be
addressed by looking at how much of the phase space of the non-dissipative
dynamics is stochastic, and seeing how much of the inspiral is spent
in the stochastic regions. The effect that passage through a stochastic region
has on gravitational wave templates can be estimated by comparing the
Lyapunov timescale $\tau_\lambda$ of these regions to their
dissipative timescale $\tau_d$\cite{foot2}. If $\tau_\lambda \ll \tau_d$, then
the unpredictability will be pronounced and gravitational waveforms for
nearby trajectories will evolve very differently, as pointed out in
Refs.\cite{us,janna}. If passage through a stochastic region happened to
coincide with passage through the LIGO frequency range, the gravitational
wave signal would be very difficult to extract.

\bigskip\noindent
Neil J. Cornish\\
\indent Department of Physics\\
\indent Montana State University\\
\indent e-mail: cornish@physics.montana.edu\\ \\



\begin{references}
\bibitem{foot2} Note, while $\tau_\lambda$ and $\tau_d$ are both coordinate
dependent, their ratio is a coordinate independent quantity.
\end{references}
\end{document}